\documentclass[11pt,a4paper]{article}   
\usepackage{fourier} 
\usepackage{amsmath,amsbsy,bm,amssymb,enumerate,mathrsfs}

\usepackage{todonotes}

\usepackage{graphicx} 
\usepackage[utf8]{inputenc} 
\usepackage{utf8math}
\usepackage{anysize}

\usepackage{subfigure}

\usetikzlibrary{arrows}
\definecolor{myred}{RGB}{163, 51, 61}
\definecolor{myblue}{RGB}{90, 107, 127}
\definecolor{mygreen}{RGB}{133, 177, 168}

\usepackage{algorithm}
\usepackage[noend]{algpseudocode}

\providecommand{\conv}{\mathrm{conv}} 

\providecommand{\sH}{{H}}
\providecommand{\sL}{{L}}

\newcommand {\Z}	  {\mathbb{Z}}

\newcommand {\R}	  {\mathbb{R}}

\newcommand{\ignore}[1]{}

\renewcommand{\epsilon}{\varepsilon}

\renewcommand{\bar}[1]{\overline{#1}}

\usepackage{amsthm}
\usepackage{thmtools}
\usepackage{thm-restate}
\theoremstyle{plain}
\newtheorem{theorem}{Theorem}

\newtheorem{lemma}[theorem]{Lemma}
\newtheorem{corollary}[theorem]{Corollary}
\newtheorem{proposition}[theorem]{Proposition}

\theoremstyle{definition}

\newtheorem{remark}{Remark}

\title{Block-Structured Integer and Linear Programming \\ in Strongly Polynomial and Near Linear Time}

\author{ 
  Jana Cslovjecsek\footnote{EPFL, Switzerland, \tt{jana.cslovjecsek@epfl.ch}}
  \and
  Friedrich Eisenbrand \footnote{EPFL, Switzerland, \tt{friedrich.eisenbrand@epfl.ch}}
  \and
  Christoph Hunkenschr\"oder \footnote{TU Berlin, Germany, \tt{hunkenschroeder@tu-berlin.ch}}
  \and 
  Lars Rohwedder \footnote{EPFL, Switzerland, \tt{lars.rohwedder@epfl.ch}, supported by the Swiss National Science Foundation project 200021-184656}
  \and 
  Robert Weismantel
  \footnote{ETH Zürich, Switzerland, \tt{robert.weismantel@ifor.math.ethz.ch}}
}

\begin{document}

\date{}

\maketitle

\begin{abstract}
  \noindent We consider \emph{integer} and \emph{linear programming}
  problems for which the linear constraints exhibit a (recursive)
  block-structure: The problem decomposes into independent and
  efficiently solvable sub-problems if a small number of constraints is deleted. A prominent example are \emph{$n$-fold} integer
  programming problems and their generalizations  which have received considerable attention in
  the recent literature.  The previously
  known algorithms for these problems are based on the
  \emph{augmentation framework}, a tailored integer programming
  variant of local search.

  In this paper we propose a different approach.
  Our algorithm relies
  on \emph{parametric search} and a new \emph{proximity bound}.
  We show that block-structured linear programming
  can be solved efficiently via an adaptation of a parametric search framework by Norton, Plotkin, and Tardos in combination with Megiddo's multidimensional search technique.
  This also forms a subroutine of our algorithm for the
  integer programming case by solving a strong relaxation
  of it.
  Then we show that, for any given optimal vertex solution of this relaxation,
  there is an optimal integer solution within
  $\ell_1$-distance independent of the dimension of the problem. This in turn allows us to find an optimal integer solution efficiently. 
  
  We apply our techniques to integer and linear programming with $n$-fold structure or bounded dual treedepth, two benchmark problems in this field. We obtain the first algorithms for these cases that are both near-linear in the dimension of the problem and strongly polynomial.
  Moreover, unlike the augmentation algorithms, our approach
  is highly parallelizable.
  

\end{abstract}
\pagebreak


\section{Introduction}
\label{sec:introduction-fritz}

We consider integer and linear programming problems that decompose into $n$ independent sub-problems after the removal of a small number of constraints
\begin{subequations}\label{eq:25}
  \begin{align}
    \max\; \bar{c}_1^T x^{(1)} + \cdots &+ \bar{c}_n^T x^{(n)} \label{eq:10}\\
    A_1 x^{(1)} + \cdots + A_n x^{(n)} &= b_0 \label{eq:11} \\
    x^{(i)} &\in Q_i \quad i=1,\dotsc,n.  \label{eq:16}    
  \end{align}
\end{subequations}
Here, the  $Q_i ⊆ ℝ_{≥0} ^{t_i}$ are polyhedra, the $A_i ∈ℤ^{r ×t_i}, \, i=1,\dots,n$ are integer matrices and $b ∈ℤ^r$ is an integer vector.
In the case where \eqref{eq:25} is to model an integer linear program, we also have the integrality constraint
\begin{subequations} 
  \begin{align}
    x^{(i)} &\in ℤ^{t_i} \quad i=1,\dotsc,n.  \label{eq:26}
  \end{align}
\end{subequations}
The $r$ equations~\eqref{eq:11} are the \emph{linking contraints} of the optimization problem~\eqref{eq:25}. If they are removed from \eqref{eq:25}, then the problem decomposes into $n$ independent linear or integer programming problems. In the literature, such problems are often coined integer or linear programming problems with \emph{block-structure} and algorithms to detect and leverage block-structure are an essential part of commercial solvers for integer and linear programming problems, see~\cite{martin1999integer,borndorfer1998decomposing}. 

\bigskip

The {main  contribution} of our paper is twofold.

\begin{enumerate}[i)]
\item We show how to efficiently solve the linear programming problem~\eqref{eq:25} by adapting the framework of Norton et al.~\cite{norton1992using} to the setting of block-structured problems. We leverage the inherent parallelism  by using the multidimensinoal search  technique of  Megiddo~\cite{DBLP:journals/jacm/Megiddo84,DBLP:journals/siamcomp/Dyer86,DBLP:journals/ipl/Clarkson86}  and obtain the following.
Let $T_{\max}$ be an upper bound on the running time for solving a linear programming problem $\max \{\bar{c}_i^T x^{(i)} ： x^{(i)} ∈Q_i\}$.
Then \eqref{eq:25} can be solved in parallel on $n$ processors,  where each processor carries out $2^{O(r^2)} (T_{\max} \log n)^{r+1}$ operations.
For technical reasons we make the mild assumption that the algorithm is linear, see Section~\ref{sec:prelim} for a definition.
  This result can be further refined if the individual block problems $\max \bar{c}_i^T x^{(i)}, x^{(i)} ∈Q_i$ can be efficiently solved in \emph{parallel} as well. \label{item:4} 
\item We furthermore provide the following new proximity bound for the integer programming variant of~\eqref{eq:25}. If all polyhedra $Q_i$ are integral and if $y^*$ is a vertex solution of the corresponding linear program \eqref{eq:25}, then there exists an optimal solution $z^*$ of the integer programming problem such that
  \begin{equation}
    \label{eq:9}
    \|y^* - z^*\|_1 ≤ (2rΔG+1)^{r+4}. 
  \end{equation}
  Here $Δ$ is an upper bound on the absolute value of an entry of the $A_i$ and $G$ is an upper bound on the $ℓ_1$-norm the \emph{Graver basis element} of the  matrices in a standard-form representation of the $Q_i$.
  This is a common parameter that we will elaborate in Section~\ref{sec:prelim}.
  The  new contribution is that this bound is independent of  $n$ \emph{and} $t$.  \label{item:5}
\end{enumerate}
Using (i) and (ii) we also derive an algorithm for the
integer programming problem~\eqref{eq:25}.
We first solve the continuous relaxation of~\eqref{eq:25} using (i).
Here we assume that $Q_i$, $i=1,\dotsc,n$, are integral
or otherwise we replace them by the convex hull of its integer solutions.
Then we find an optimal integer solution via dynamic programming 
using (ii) to restrict the search space.
All techniques above can be applied recursively, e.g., to solve
block-structured problems where the polytopes $Q_i$, $i=1,\dotsc,n$,
have a block-structure as well.
\subsection*{Applications}
Block-structured integer programming has been studied mainly in connection
with fpt-algorithms.
An algorithm is \emph{fixed-parameter tractable (fpt)} with respect to a parameter $k$ derived from the input, if its running time is of the form $f(k) \cdot n^{O(1)}$ for some computable function $f$.

An important example of block-structured integer programs
is the \emph{$n$-fold integer programming} problem.
In an $n$-fold integer program~\cite{de2008n}, the polyhedra $Q_i$ are given by systems $B_i x^{(i)} = b_i$, $0 \leq x^{(i)} \leq u$,
with $B_i \in \Z^{s \times t}$ all of the same dimension.
We note that in some earlier works it was assumed that $A_1 = A_2 = \cdots = A_n$ and $B_1 = B_2 = \cdots = B_n$.

Algorithms for $n$-fold integer programming has been used, for example in~\cite{knop2017scheduling,chen2017scheduling,jansen2018scheduling} to derive novel fpt-results in scheduling. 
Moreover, they have been successfully applied to derive fpt-results for string and social choice problems~\cite{KnopKM17-bribery,knop2017combinatorial}.
We refer to the related work section for a comprehensive literature review
on algorithmic results for $n$-fold integer programming. We  emphasize three main applications of the present article. 

\begin{enumerate}[a)]
\item We obtain a strongly polynomial and nearly linear time algorithm for
 $n$-fold integer programming. Our algorithm requires
      \begin{equation*}
        2^{O(r s^2)} (rs\Delta)^{O(r^2 s + s^2)} (nt)^{1 + o(1)}
      \end{equation*}
      arithmetic operations, or alternatively $2^{O(r^2 + r s^2)} \log^{O(rs)} ( nt \Delta)$ parallel operations
      on $(rs\Delta)^{O(r^2 s + s^2)} nt$ processors.
      Previous algorithms in the literature either had at least a quadratic (and probably higher)
      dependence on $nt$ or an additional  factor of $\Phi$, which is the
      encoding size of the largest integer in the input.
      Moreover, our algorithm is the first parallel algorithm.
\item We present an algorithm for
 $n$-fold linear programming.
 Our algorithm requires
      \begin{equation*}
        2^{O(r^2 + rs^2)} (nt)^{1 + o(1)}
      \end{equation*}
      arithmetic operations, or alternatively $2^{O(r^2 + r s^2)} \log^{O(rs)} (nt)$ parallel operations on $nt$ processors.
      This extends the class of linear programs known to be solvable in strongly
      polynomial time and those known to be parallelizable.
\item Another parameter that gained importance for integer programming is the (dual) \emph{treedepth} of the constraint matrix, which can be defined recursively as follows.
The empty constraint matrix has treedepth $0$; a matrix with treedepth $d > 0$ is either block-diagonal and the maximal treedepth of each block is $d$, or there exists a row such that deleting this row leads to a matrix of treedepth $d-1$.
	In terms of treedepth, we also obtain a strongly polynomial algorithm.
	More precisely, our algorithm requires 
\begin{equation*}
	2^{O(d 2^d)} \Delta^{O(2^d)} h^{1 + o(1)}
\end{equation*}
	arithmetic operations, or alternatively $2^{O(d 2^d)} \log^{O(2^d)}(\Delta h)$ parallel operations on $\Delta^{O(2^d)}h$ processors.
    Here $h$ is the number of variables.
    Furthermore, we present a running time analysis
    for the corresponding linear programming case.
\end{enumerate} 

\subsection*{Related work}
Integer programming can be solved in polynomial time, if the dimension is fixed~\cite{kannan1987minkowski,lenstra1983integer}.
Closely related to the results that are presented here are dynamic programming approaches to integer programming~\cite{papadimitriou1981complexity}. In~\cite{DBLP:journals/talg/EisenbrandW20}  it was shown that an integer program $\max\{c^Tx : Ax = b, \, l ≤ x ≤ u, \,  x ∈ ℤ^n\}$ with $A ∈ ℤ^{s ×n}$  can be solved in time $(sΔ)^{O(s^2)} n$ and in time $(sΔ)^{O(s)}$ if there are no upper bounds on the variables.
Jansen and Rohwedder~\cite{jansen2018integer} obtained better constants in the exponent of the running time of integer programs without upper bounds. Assuming the exponential-time hypothesis, a tight lower bound was presented by Knop et al.~\cite{knop2019tight}.

The first fixed-parameter tractable algorithm for $n$-fold integer programming
is due to Hemmecke et al.~\cite{hemmecke2013n-fold} and is with respect to
parameters $\Delta,r,s$ and $t$.
Their running time is $O(n^3 \phi)  \Delta^{O(t(rs+st))}$ 
where $\phi$ is the encoding size of the largest absolute value of a component of  the input.
  
The exponential dependence on $t$ was removed by Eisenbrand et al.~\cite{eisenbrand2018faster} and Kouteck\'y et al.~\cite{koutecky2018parameterized}.
The first strongly polynomial algorithm for $n$-fold integer programs was given by Kouteck\'y et al.~\cite{koutecky2018parameterized}.
The currently fastest algorithms for $n$-fold integer programming are provided by Jansen, Lassota and Rohwedder~\cite{jansen2019near} and Eisenbrand et al.~\cite{eisenbrand2019algorithmic}.
While the first work has a slightly better parameter-dependency, the second work achieves a better dependency on the number of variables.
We note that the results above are based on an augmentation framework,
which differs significantly from our methods.
In this framework an algorithm iteratively 
augments a solution ultimately converging to the optimum.
This requires $\Omega(n)$ sequential iterations,
which makes parallelization hopeless.

Other variants of recursively defined block structured integer programming
problems were also considered in the literature.
Notable cases include the tree-fold integer programming
problem introduced by Chen and Marx~\cite{DBLP:conf/soda/ChenM18}.
This case is closely related to dual treedepth and can be analyzed in a similar
way using our theorems.
The currently best algorithms for treedepth obtain a running time of $\Delta^{O(d 2^d)} \Phi^2 n^{1 + o(1)}$~\cite{eisenbrand2019algorithmic}.

For our algorithm it is
important to use an LP relaxation where $Q_i$, $i=1,\dotsc,n$, are integral.
For many settings (e.g., for the $n$-fold case) this is stronger than
using a naive LP relaxation.
The idea of using a stronger relaxation than the naive LP relaxation was 
also used in~\cite{MIMO}, where the authors consider a 
high-multiplicity setting.
Roughly speaking, if there are only a few 
types of polyhedra $Q_i$ that might be repeated several times, the authors also 
obtain a proximity result independent of the dimension. However, this 
proximity result still depends on the number of variables $t$ per 
block, and the number $\tau$ of different polyhedra $Q_i$.
%

\section{Preliminaries}
\label{sec:prelim}
We now explain the notions of a \emph{Graver basis}
and of a \emph{linear algorithm}. 
The former definition is used in our proximity bound and
will play an important role in the proof. The latter
is a property of some algorithms which will be central in our algorithm
for block-structured linear programming problems.
\subsubsection*{Graver basis}
Two vectors $u,v ∈ ℝ^n$ are said to be \emph{sign compatible} if $u_i ⋅ v_i ≥0$ for each $i$.
Let $C ∈ℤ^{m ×n}$ be an integer matrix.
An integral element $u ∈ \ker(C) ∩ ℤ^n$  is \emph{indecomposable} if it is not the sum of two other sign-compatible non-zero integral elements of   $\ker(C) $. The set of indecomposable and integral elements from the kernel of $C$ is called the \emph{Graver basis} of $C$~\cite{graver1975foundations}, see also \cite{onn2010nonlinear,loera2012algebraic}.
Each integral polyhedron $Q ⊆ ℝ_{≥0}^{t}$ is the integer hull of a rational polyhedron $P = \{x ： Ax ≤b \}$, where $A ∈ℤ^{m ×t}$ and $b ∈ ℤ^m$.
We assume that $Q$ is given implicitly as the integer hull of a polytope $P$, where $P$ is explicitly given in equality form.
from the description above, this representation of $P$ can be achieved by splitting $x$ into positive and negative variables, and adding slack variables, this is $Ax^+ - Ax^- + Iz = b$, $x^+,x^-,z \geq 0$, where $I$ is the identity matrix of suitable dimension.
\subsubsection*{Linear algorithms}
\label{sec:linear-algorithms-1}

The concept of accessing numbers in the input of an algorithm by \emph{linear queries} only is a common feature of many algorithms and crucial in this paper as well. Let $λ_1,\dots,λ_k$  be numbers in the input of an algorithm. The algorithm is \emph{linear} in this part of the input if it does not query the value of these numbers, but queries linear comparisons instead. This means that the algorithm can generate numbers $a_1,\dots,a_k∈ℝ$ and $β∈ℝ$ and queries whether
\begin{equation}
  \label{eq:27}
  a_1 λ_1 + \cdots +  a_k λ_k  ≤ β
\end{equation}
holds. Such a query is to be  understood as a call to an oracle that is not implemented in the algorithm. 

Many sorting algorithms such as \emph{quicksort} or \emph{merge sort} are linear in the input numbers $λ_1,\dots,λ_k$ to be sorted. Each basic course in algorithms treats lower bounds of linear sorting algorithms for example, see, e.g.~\cite{CLRS2001}.

This feature of linearity  is also common in discrete optimization. 
For example, the  well known \emph{simplex algorithm} for linear programming  $\max\{ c^Tx ： Ax≤ b\}$ is linear in $b$ and in $c$. 
Let us assume that $A ∈ ℝ^{m×n}$  is of full column rank and now 
let us convince ourselves that, in order to test feasibility and optimality of a basis $B ⊆ \{1,\dots,m\}$ one only needs linear queries on $c$ and $b$.
Recall that a basis is a set of $n$ indices corresponding to linearly independent rows of $A$. 
The basis is feasible if $A (A_B^{-1}b_B) ≤ b$ holds. These are  $m$ linear queries involving the components of  $b$.  Similarly, $B$ is an optimal basis if $c^T A_B^{-1}≥ 0$ holds and this corresponds to $n$ linear queries involving the components of  $c$. 

\medskip
\noindent 
We use the following {\bf notation}. If an algorithm is linear in some part of the numbers in its input $λ$, then we denote this part with a bar, i.e., we write $\bar{λ}$. Thus in the case of the simplex algorithm, we would write that it solves a linear program $\max\{ \bar{c}^Tx ： Ax≤ \bar{b}\}$ to indicate, which input numbers are only accessible via linear queries. In \emph{our setting}, i.e. in \eqref{eq:25}, we assume that the objective function vector $\bar{c}$  is only accessible via linear queries. Our algorithm thus will be \emph{linear} in $\bar{c}$.

\subsubsection*{Parallel algorithms}
A central theme in this paper is parallization. The model we are interested in
is the \emph{parallel RAM (PRAM)}, in which a fixed number of processors
have access to the same shared memory. More precisely, we consider
the \emph{concurrent read, concurrent write (CRCW)} variant. In this model we are
typically interested in algorithms that run in polylogarithmic time
on a polynomial number of processors.
We refer the reader to~\cite{DBLP:books/daglib/0031448} for further details on the model.

\section{Solving the LP by Parametric Search and Parallelization}
\label{sec:parametric}
We now describe how to solve the linear programming problem~\eqref{eq:25} efficiently and in parallel. The method that we lay out is based on a technique of Norton, Plotkin and Tardos~\cite{norton1992using}.
The authors of this paper show the following. Suppose there is an algorithm that solves a linear programming problem in time $T(n)$, where $n$ is some measure of the length of the input and let us suppose that we change this linear program by adding $r$ additional constraints.
Norton et al.\ show that this augmented linear program can be solved in time $(T(n))^{r+1}$. 
A straightforward application of this technique to our setting would yield the following.
We interpret the starting linear program as the problem~\eqref{eq:25} that is obtained by deleting the $r$ linking constraints.
This problem is solved in time $Ω(n)$ by solving the $n$ individual linear programming problems $\max\{ \bar{c}_i^T x ： x ∈ Q_i\}$.
Using the result~\cite{norton1992using} out of the box would yield a running time bound of $Ω(n^{r+1})$. The main result of this section  however is the following. 

\begin{theorem}
\label{thr:lp-parallel}
Suppose there  are  algorithms that solve   $\max \{\bar{c}_i^Tx ： x ∈  Q_i\}$ on $R_i$ processors using at most $T_{\max}$ operations on each processor and suppose that these algorithms are linear in $\bar{c}_i$.  Then there is an algorithm that  solves the linear programming problem~\eqref{eq:25} 
on $R=\sum_{i} R_i$ processors that requires  $2^{O(r^2)}(T_{\max}\log(R))^{r+1}$ operations on each processor. This algorithm is linear in $\bar{c}$.
\end{theorem}

\begin{remark}
  \label{rem:1} 
  Obviously, the problems $\max \{\bar{c}_i^Tx ： x ∈  Q_i\}$ can be solved independently in parallel. If $T_{\max}$ is an upper bound on the running times of these algorithms, then Theorem~\ref{thr:lp-parallel} provides a sequential running time of
  \begin{displaymath}
    2^{O(r^2)} n ⋅ (T_{\max} \log n)^{r+1} =  2^{O(r^2)} n^{1 + o(1)}  ⋅ T_{\max}^{r+1}
  \end{displaymath}
  to solve the block-structured linear programming problem~\eqref{eq:25}. Theorem~\ref{thr:lp-parallel} is stated in greater generality in order to use the potential of parallel algorithms that solve the  problems  $\max \{\bar{c}_i^Tx ： x ∈  Q_i\}$  themselves. This makes way for a refined analysis of linear programming problems that are in recursive block structure as demonstrated in our applications. 
\end{remark}


 \emph{The novel elements of this chapter} are the following. 
    We leverage the massive parallelism that is exhibited in the solution of the Lagrange dual in the framework of Norton et al.
    We furthermore  present an analysis of the multidimensional search
    technique of Megiddo in the framework of linear algorithms that
    clarifies that this technique can be used in our setting. Roughly speaking, multidimensional search deals with the following problem. Given $m$ hyperplanes $a_i^T λ = \bar{f}_i, i=1,\dots,m$ and $\bar{λ} ∈ℝ^r$, the task is to understand the orientation of $\bar{λ}$ w.r.t. each hyperplane.
    Megiddo shows how to do this in time $2^{O(r)} ⋅ m \log^2(m)$ while   the total  number of  linear queries involving $\bar\lambda$ is bounded by $2^{O(r)}\log(m)$. This is crucial for us  and  implicit in his analysis.
    We will make this  explicit in the appendix of this paper.
    Finally, the algorithm itself is linear in $\bar{c}$.
    In~\cite{norton1992using} it is not immediately obvious that linearity
    can be preserved.
    This is important for problems in recursive block structure. 


\subsection{The technique of Norton et al.}
\label{sec:fram-nort-plotk}

We now explain the algorithm to solve the linear program~\eqref{eq:25}.  In the following we write
\begin{align*}
  A & = (A_1 \cdots A_n) \in \R^{r \times (t_1 + \cdots + t_n)}, \\
  x^T  &=  ( {x^{(1)}}^T  \cdots   {x^{(n)}}^T) \in\R^{t_1 + \cdots + t_n}, \text{ and } \\
  \bar{c}^T  &=  (\bar{c}_1^T  \cdots  \bar{c}_n^T  ) \in\Z^{t_1 + \cdots + t_n}. 
\end{align*}

It is well known that a linear program can be solved via \emph{Lagrangian relaxation}. We refer to standard textbooks in optimization for further details, see, e.g.~\cite{MR2061575,Schrijver86}.  
By dualizing the linking constraints~\eqref{eq:11} the Lagrangian $L(λ)$ with weight $λ ∈ℝ^r$ is the 
linear programming problem
\begin{equation}
  \label{eq:13}
  \begin{split}
    L(λ) =   \max\ \bar{c}^T x &- \lambda (Ax - b) \\
    x^{(i)} &\in Q_i \quad i=1,\dotsc,n.
  \end{split}
\end{equation}

The Lagrangian dual is the 
task to solve the convex optimization problem 
\begin{equation}
  \label{eq:7}
  \min_{\lambda\in\R^r} L(\lambda). 
\end{equation}

For a given ${\bar{λ}}$ the value of $L(\bar{λ})$ can be found by solving the independent optimization problems
\begin{equation}
  \label{eq:29}
  \max\{ (\bar{c}_i^T - \bar{λ}^TA_i) x^{(i)} ：x^{(i)} ∈Q_i \}
\end{equation}
with the corresponding algorithms that are linear in this objective function vector.
The objective function vector of \eqref{eq:29} is $\bar{c}_i - A_i^T \bar{λ}$. A query on    this objective function vector that is posed by the algorithm that solves~\eqref{eq:29}  is thus of the form 
\begin{equation}
  \label{eq:30}
  a^T  \bar{λ} ≤ \bar{f}. 
\end{equation}
Here $\bar{f}$ is an affine  function in the components of $\bar{c}_i$ defined by the query.

\medskip
We are now ready to describe the main idea of the framework of Norton et al.
Assume that we have an algorithm $\mathscr{A}_k$  that solves the restricted Lagrangian dual 
\begin{equation}
  \label{eq:8}  
  \min_{λ ∈S} L(\lambda),
\end{equation}
where $S$ is any  $k$-dimensional affine subspace of $\R^r$ defined
by $S = \{\lambda\in\R^r : D\lambda = \bar d \}$ for
a matrix $D\in\R^{(r - k) \times r}$ of $r - k$ linear independent
rows and $\bar d \in \R^{r - k}$.
To be precise, $\mathscr{A}_k$ is delivering the optimal solution $\bar{λ}_S$ of~\eqref{eq:8} (as an affine function in $\bar c$) as well as an optimal solution $x^*∈ Q_1 × \cdots ×Q_n$ of the linear program $L(\bar{λ}_S)$. 
We assume that this algorithm is linear in 
$\bar c$ and that $\bar d$  is a vector with each  component   being a linear function in $\bar{c}$. In other words, the restricted Lagrangian stems from constraining the $λ$ to satisfy $n-k$ linear independent queries of the form~\eqref{eq:30} with equality.

The algorithm $\mathscr{A}_0$ simply returns the unique optimal solution $D^{-1}\bar{d}$ which is an affine  function in $\bar{c}$ together with an optimal solution of the corresponding linear program~\eqref{eq:13}. An algorithm $\mathscr{A}_r$ is an algorithm for the unrestricted version of the Lagrangian dual \eqref{eq:7}.

We now describe how to construct the algorithm $\mathscr{A}_{k+1}$ with an algorithm $\mathscr{A}_k$ at hand.     
To this end, let $S = \{\lambda\in\R^r : D\lambda = \bar d \}$ be of dimension $k+1$, i.e., we assume that $D$ consists of $r-k-1$ linear independent rows.
Furthermore,  let $\bar{λ}_S ∈S$ be an optimal solution of the restricted dual~\eqref{eq:8} that is unknown to us.

Let us denote the algorithm that evaluates the Lagrangian $L(\bar{λ})$ by $ℰ$.  
The idea is now to run $ℰ$ on  $L(\bar{λ}_S)$. We can do this, even though $\bar{λ}_S$ is still unknown,  as long as we answer each linear  query~\eqref{eq:30} that occurs in $ℰ$  as if it was queried for $\bar{λ}_S ∈S$.  The point $\bar{λ}_S$  is then any point that satisfies all the queries.

Let $a^T \bar{λ} ≤\bar{f}$ be a query~\eqref{eq:30}. Clearly, if $a$ is in the span of rows of $D$, then the query can be answered by a linear query on the right hand sides, i.e., on $\bar{c}$. We thus assume that $a$ is not in the span of the rows of $D$. 
We next show that, by three calls on the algorithm $\mathscr{A}_k$ we can decide whether 
\begin{enumerate}[(i)]
\item  $a^T \lambda^* = \bar f$ for some optimal point $\lambda^*\in S$,\label{item:6}
\item $a^T \lambda^* > \bar f$, or\label{item:7}
\item  $a^T \lambda^* < \bar f$ for each optimal solution $λ^*$\label{item:8}
\end{enumerate}
which means that we can answer the query as if it was asked for $\bar{λ}_S$. 
Let $ \epsilon > 0$ be an unspecified very small value. 
We use $\mathscr{A}_k$ to find optimal solutions  $\bar{λ}_L, \bar{λ}_R$ and $\bar{λ}_0$ of the Lagrangian that is restricted to 
\begin{align*}
  S_L &  = S ∩ \{\lambda : a^T λ = \bar{f} - ε \} \\
  S_0 &  = S ∩ \{\lambda : a^T λ = \bar{f} \} \\ 
  S_R &  = S ∩ \{\lambda : a^T λ = \bar{f} + ε \} 
\end{align*}
respectively. Since $\bar{λ}_L, \bar{λ}_R$ and $\bar{λ}_0$ are affine functions in $\bar{c}$, we can compare their corresponding values and since $L(λ)$ is convex, these comparisons allow us to decide whether~(\ref{item:6}), (\ref{item:7}) or (\ref{item:8}) holds.

%

The value $ \epsilon$ does not have to be provided explicitly.
It can be treated symbolically. For instance, when we recurse on $S_L$,
we have to answer a series of comparisons of the form
$u^T \bar c  +  y  \epsilon \le z$. If $y$ is positive, this is the query
$u^T \bar c  <  z$ which can be answered by querying $u^T \bar c  ≤  z$ and $u^T \bar c  ≥  z$.

We have shown how to simulate the algorithm that computes $L(λ)$ as if it was on input $\bar{λ}_S$. It remains to describe how to retrieve $\bar{λ}_S$ as a linear function in $\bar{c}$. This is done as follows. Let $x^*$ be an optimal solution that has been found by the above simulation and 
let $U λ = \bar{u}$ be the system of equations that is formed by setting all queries that have been answered according to (\ref{item:4}).  The value of $L(\bar{λ}_S)$ is equal to
\begin{displaymath}
  \max\left\{  \bar c^T x^* - \lambda^T (Ax^* - b) ： \lambda\in\R^r,  \begin{pmatrix} D \\ U \end{pmatrix} \lambda = \begin{pmatrix} \bar{d} \\ \bar{u} \end{pmatrix} \right\} .
\end{displaymath}
Thus, if $(Ax^*-b)^T$ can be expressed as a linear combination of the rows in $D$ and $U$,
then any point in the subspace above is optimal and it can be found with Gaussian elimination. 
Otherwise, the Lagrange dual restricted to $S$ is unbounded.

We analyze the running time of the algorithm $\mathscr{A}_{k+1}$. Let $T$ be the running time of the algorithm that evaluates $L(λ)$ where a linear query \eqref{eq:30} counts as one. Then, the running time of $A_{k+1}$ is $3 ⋅ T$  times the running time of the algorithm $A_{k}$. This shows that the running time of $\mathscr{A}_r$ is bounded by $(3T)^{r+1}$.

\subsection{Acceleration by Parallelization and Multidimensional Search} 
\label{sec:mult-search}

The block structured linear program~\eqref{eq:25} has the important feature that, for a given $\bar{λ} ∈ ℝ^r$, the value of $L(\bar{λ})$ can be computed by solving the $n$  linear programming problems~\eqref{eq:29} in parallel.  We now explain how to exploit this and prove Theorem~\ref{thr:lp-parallel}.

For the sake of a more accessible treatment, let us assume for now that each of these problems can be solved in time $T_{\max}$ on an individual processor. This means that the evaluation of $L(λ)$ can  be carried out with algorithm $ℰ$ on $n$ processors by algorithms that are linear in their respective objective function vectors.  We are now looking again at the construction of the algorithm $\mathscr{A}_{k+1}$ with an algorithm $\mathscr{A}_{k}$ and the algorithm $ℰ$  at hand.

In one step of the parallel algorithm $ℰ$, there are at most $n$ queries of the form~\eqref{eq:30} that come from the individual sub-problems $Q_i$, $i=1,\dots,n$. In the previous paragraph, we were answering these queries one-by-one according to $\bar{λ}_S$ by calling three times $\mathscr{A}_k$. We can save massively by using  Megiddo's multidimensional search technique
and Clarkson and Dyer's improvement~\cite{DBLP:journals/jacm/Megiddo84,DBLP:journals/siamcomp/Dyer86,DBLP:journals/ipl/Clarkson86}.

\begin{restatable}[Megiddo]{theorem}{megiddo}
   \label{prop:megiddo-pram}
  Let $\bar\lambda\in\R^r$ and 
  consider a set of $m$
  hyperplanes
  \begin{equation*}
    H_i = \{\lambda\in\R^r :  a_i^T \lambda = \bar{f}_i\} , \,    i=1,\dotsc,m.
  \end{equation*}
  There is an algorithm that
  determines for each hyperplane whether 
  $a_i^T \bar\lambda = \bar{f}_i$,
  $a_i^T \bar\lambda < \bar{f}_i$,
  or $a_i^T \bar\lambda > \bar{f}_i$
  in $2^{O(r)}\log^2(m)$ operations on $O(m)$ processors. 
  Moreover, the total (sequential) number of comparisons dependent on $\bar\lambda$
  is at most $2^{O(r)}\log(m)$.
\end{restatable}
  

 The statement of Theorem~\ref{prop:megiddo-pram} is in the framework of linear algorithms. The proof is implicit in the papers~\cite{DBLP:journals/jacm/Megiddo84,DBLP:journals/siamcomp/Dyer86,DBLP:journals/ipl/Clarkson86}. We nevertheless provide a proof in the appendix. To get an intuition on why the  number of queries involving $\bar{λ}$ is this low,  we explain here why the Theorem is true in the base-case  $r = 1$.

To this end, assume that $a_i \neq 0$ for all $i$ and 
  compute the median  $\bar M$  of the numbers $\bar{f}_i / a_i$. 
  Then, for each hyperplane $H_i$ one checks  whether $\bar{f}_i / a_i \le \bar M$ and $\bar{f}_i / a_i \ge \bar M$ holds.
  The median computation requires $O(\log(m))$ operations on $O(m)$ processors,
  for example by a straightforward implementation of merge sort, see~\cite{DBLP:books/daglib/0031448}.
  Now compare $\bar{\lambda} \le \bar M$ and $\bar{\lambda} \ge \bar M$. From the result we can derive an answer
  for $m/2$ of the hyperplanes. Thus the  total number of linear queries involving $\bar{λ}$  is bounded by $O(\log m)$.

\begin{figure}
  \centering

\begin{tikzpicture}
  \useasboundingbox (-1, 0) rectangle (11.5, 5.5);

  \draw[thick, -triangle 45] (0, 4.5) -- (0, 2.5) node[pos=0.5, left] {time};

  \draw[thick] (1, 1.5) rectangle (3, 5.5);
  \draw[thick, dashed] (1, 2.3) -- (3, 2.3);
  \draw[thick] (1, 3.1) -- (3, 3.1);
  \draw[thick] (1, 3.9) -- (3, 3.9);
  \draw[thick, dashed] (1, 4.7) -- (3, 4.7);
  \node at (2, 1) {$P_1$};

  \draw[thick] (3.5, 1.5) rectangle (5.5, 5.5);
  \draw[thick, dashed] (3.5, 2.3) -- (5.5, 2.3);
  \draw[thick] (3.5, 3.1) -- (5.5, 3.1);
  \draw[thick] (3.5, 3.9) -- (5.5, 3.9);
  \draw[thick, dashed] (3.5, 4.7) -- (5.5, 4.7);
  \node at (4.5, 1) {$P_2$};

  \node at (6.5, 3.5) {$\cdots$};
  \node at (6.5, 1) {$\cdots$};

  \draw[thick] (7.5, 1.5) rectangle (9.5, 5.5);
  \draw[thick, dashed] (7.5, 2.3) -- (9.5, 2.3);
  \draw[thick] (7.5, 3.1) -- (9.5, 3.1);
  \draw[thick] (7.5, 3.9) -- (9.5, 3.9);
  \draw[thick, dashed] (7.5, 4.7) -- (9.5, 4.7);
  \node at (8.5, 1) {$P_n$};

  \node at (2, 3.5) {$a_1^T \bar\lambda_S \le \bar f_1$};
  \node at (4.5, 3.5) {$a_2^T \bar\lambda_S \le \bar f_2$};
  \node at (8.5, 3.5) {$a_n^T \bar\lambda_S \le \bar f_n$};

  \draw[thick, mygreen] (0.8, 2.9) rectangle (9.7, 4.1);
\end{tikzpicture}

  \caption{A parallel run of algorithm $ℰ$ }
\label{fig:2}
\end{figure}
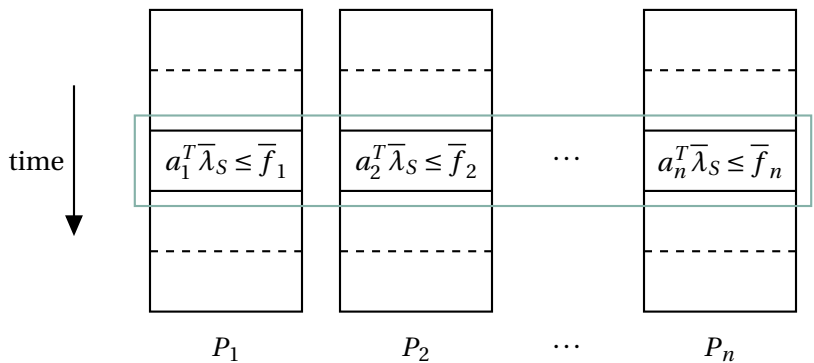

 \medskip  
  Back to the algorithm $ℰ$ that evaluates the Lagrangean at $\bar{λ}_S$. This algorithm runs the optimization problems over the $Q_i$ in parallel.  At a given time step, these $n$ algorithms make queries of the form~\eqref{eq:30}
  \begin{displaymath}
    a_1^T \bar{λ}_S ≤ \bar{f}_1, \cdots,  a_n^T \bar{λ}_S ≤ \bar{f}_n,
  \end{displaymath}
  see Figure~\ref{fig:2}. 
  Recall that the $\bar{f}_i$ are affine functions of $\bar{c}$. 
  We answer these queries with Megiddo's algorithm in time $2^{O(r)} \log^2 n$ operations on $n$ processors and a total number of  $2^{O(r)}  \log n$ linear comparisons on $\bar{λ}_S$. These linear queries  can be answered by $3$ calls to algorithm $\mathscr{A}_k$ as we explained in the previous section. If $T_{\max}$ is the maximum running time of the parallel algorithms in  $ℰ$, then the running time of $\mathscr{A}_{k+1}$ is now $2^{O(r)} \log^2 n T_{\max} $ plus $ 2^{O(r)}  \log n \,  T_{\max}$  times the running time of $\mathscr{A}_k$  on $n$ processors. This sows that the running time of $\mathscr{A}_r$  is bounded by $2^{O(r^2)}    (T_{\max} \log n)^{r+1}$. This almost completes  the proof of Theorem~\ref{thr:lp-parallel} for the case that each algorithm for the optimization problems over $Q_i$  is sequential.   What is missing is how to find the optimal  solution $x^*$ of~\eqref{eq:25}.  We explain this now.



  \begin{proof}[Proof of Theorem~\ref{thr:lp-parallel}]

    By following the lines of the argument above, but this time
    assuming that the linear optimization problems
    $\max \{\bar{c}_i^Tx ： x ∈ Q_i\}$ are solved on $R_i$  processors using at
    most $T_{\max}$ operations on each processor, we obtain that the
    algorithm $\mathscr{A}_r$ requires
    $2^{O(r^2)}(T_{\max}\log(R))^{r+1}$ operations on
    $R = ∑_{i=1}^nR_i $ processors. The optimal solution
    $\bar{λ}_{OPT}$ of the Lagrangian~\eqref{eq:7} can thus be found
    in this time bound. It remains to show how to find an optimal
    solution $x^*$ of the linear program~\eqref{eq:25}. The algorithm
    $\mathscr{A}_r$ proceeds by various calls to the algorithm $ℰ$
    that, in the course of $\mathscr{A}_r$ finds vertices
    $v_1,\dots,v_ℓ$ of $Q_1×\cdots ×Q_n$.  The value of the
    Lagrangian~\eqref{eq:7} and the value of the Lagrangian in which
    $Q_1 × \cdots ×Q_n$ is replaced by the convex hull
    $\conv\{v_1,\dots,v_ℓ\}$ of these vertices is the same.    
    Therefore, an optimal solution of the
    LP~\eqref{eq:25} can be found by restricting to the convex hull of
    these vertices. 
    This is the linear program
\begin{align*}
  \max  ∑_{i=1}^ℓ  (\bar{c}^T v_i) \,  μ_i  &\\
  A\left( ∑_{i=1}^ℓ μ_i v_i \right) & = b \\
  ∑_{i=1}^ℓ μ_i & = 1 \\
  μ≥0.          
\end{align*}
The dual of this linear program is a linear program in dimension $r+1$  with $ℓ$ constraints and this can be solved in time $2^{O(r^2)} ⋅ ℓ $ with Megiddo's algorithm. We now argue that the number $ℓ$ of vertices is bounded by $2^{O(r^2)} (T_{\max} \log R)^r$ which in turn implies that this additional work can be done on one processor and the claimed  running time bound holds.

The algorithm $\mathscr{A}_{k+1}$  runs $ℰ$ and makes $2^{O(r)} T_{\max} \log R$ calls to the algorithm $\mathscr{A}_k$. 
This shows that the total number of calls to $ℰ$ that are incurred by the algorithm $\mathscr{A}_r$  is bounded by $2^{O(r^2)} (T_{\max} \log R)^r$ and bounds the number of vertices as claimed. This finishes the proof of the main result of this section. 
\end{proof}

\section{Proximity}
\label{sec:proximity}
Our goal is to describe a relaxation of the block-structured integer programming problem \eqref{eq:25}, that can be solved efficiently with the techniques of Section \ref{sec:parametric} and has LP/IP proximity $(r\Delta G)^{O(r)}$.
The following proposition shows that the standard relaxation, obtained by relaxing the integrality constraints $x^{(i)} \in \Z^{t_i}$ to $x^{(i)} \in \R^{t_i}$ for $i=1,\dots,n$, is not sufficient.

\begin{restatable}{proposition}{lprelbad}
	\label{prop:1}
	There exists a family of block-structured integer programming problems such that the $\ell_∞$-distance of an optimal solution $x^*$  of the standard LP-relaxation to each integer optimal solution $z^*$  is bounded from below by
	\begin{displaymath}
	\|x^*-z^*\|_\infty  = Ω(n). 
	\end{displaymath}
\end{restatable}
The family demonstrating this is discussed in the appendix.
\medskip

We are interested in a relaxation which is closer to the optimal integer solution. Using the block-structure of the problem, we replace the polyhedra $Q_i$ by their integer hull $(Q_i)_I$, removing only fractional solutions but no integer solutions of the standard relaxation.
Note that this strengthened relaxation is still in the block-structured setting of~\eqref{eq:25} and we can suppose that the $Q_i$ are integral polyhedra. 

We now show that for such a block-structured programming problem~\eqref{eq:25} with integral polyhedra $Q_i$, an optimal vertex solution of its relaxation is close to an optimal solution of the integer problem. More precisely, we prove the following result.

\begin{theorem}
  \label{thr:3}
  Let $x^*\in \R^{t_1+\dots+t_n}$ be an optimal vertex solution of the relaxation~\eqref{eq:25} with integral polyhedra $Q_i$. Moreover, assume that $G$ is an upper bound on every Graver basis element of the matrices in standard form representation of the $Q_i$. There exists an optimal integer solution $z^*\in \Z^{t_1+\dots+t_n}$ of~\eqref{eq:25} with
  \begin{displaymath}
    \| x^* - z^*\|_1 ≤ (r \Delta G)^{O(r)}. 
  \end{displaymath}
\end{theorem}
Theorem \ref{thr:3} shows that the proximity bound for $\| x^* - z^*\|_1$ does neither depend on the number of blocks $n$ nor on the numbers $t_i$ that is the ambient dimension of the block polyhedra $Q_i$, using known bounds on the Graver basis, see e.g.~\cite{DBLP:journals/talg/EisenbrandW20}. 

\vspace{.5cm}

\begin{figure}
  \centering
  \includegraphics[scale=1]{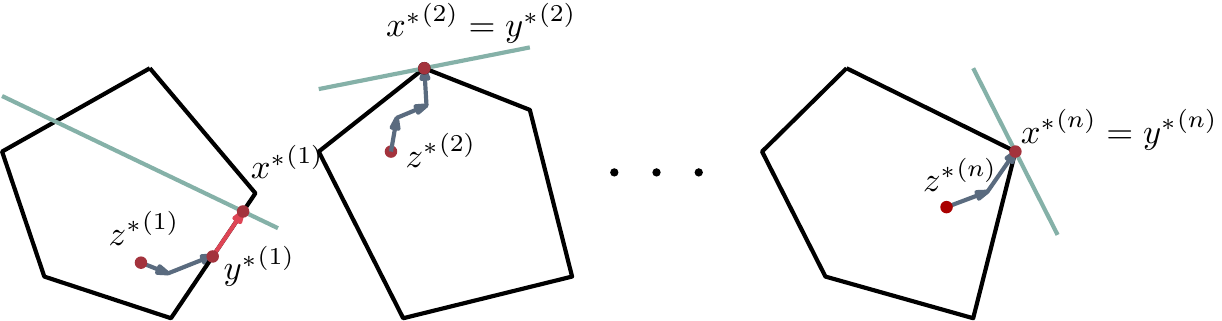}
  \caption{The proof of Theorem~\ref{thr:3}.}
  \label{fig:1}
\end{figure}

Next we present an {\bf overview of the proof}, see Figure~\ref{fig:1}. Throughout this section, we assume that the polyhedra $Q_i$ are integral, $x^*$ is an optimal vertex solution of the relaxation~\eqref{eq:25} that we partition into its blocks ${x^* }^{(i)} \in \R^{t_i}$ for $i=1,\dots,n$. 
\begin{itemize}\itemsep0em 
\item We let ${y^*}^{(i)}$ be a nearest integer point to ${x^*}^{(i)}$ with respect to the $\ell_1$-norm that lies on the minimal face of $Q_i$ containing ${x^*}^{(i)}$. In Proposition~\ref{thr:4} we show the bound $\|{x^*}^{(i)} - {y^*}^{(i)}\|_1 ≤ r^2 G$.
\item We let $z^*$ be an optimal integer solution such that $\|y^* - z^*\|_1$ is minimal. In Theorem~\ref{thr:5} we show $\|y^* - z^* \|_1 ≤ (r \Delta G)^{O(r)}$.
\item Since at most $r$ of the ${x^*}^{(i)}$ are non-integral and in particular not equal to ${y^*}^{(i)}$ (Lemma~\ref{lem:bound-fractionality}), the theorem follows by applying the above bounds
\begin{align*}
  \| x^* - z^* \|_1 & ≤  \| x^* - y^* \|_1 +  \| y^* - z^* \|_1 \\
                    & ≤  r^3 G + (r \Delta G)^{O(r)} \\
                    & ≤  (r \Delta G)^{O(r)}. 
\end{align*}
\end{itemize}

\begin{lemma}
	\label{lem:bound-fractionality}
	Let $({x^*}^{(1)},\dots,{x^*}^{(n)})$ be a vertex solution of the relaxation~\eqref{eq:25} with integral polyhedra $Q_i$. All but $r$ of the ${x^*}^{(i)}$ are vertices of the $Q_i$ respectively and thus all but $r$ of the ${x^*}^{(i)}$ are integral.
\end{lemma}

\begin{proof}
	For the sake of contradiction, suppose that ${x^*}^{(1)},\dots, {x^*}^{(r+1)}$ are not vertices of the respective $Q_i$. Then, for each $i=1,\dots,r+1$ there exists a non-zero vector $d_i \in \R^{t_i}$ such that ${x^*}^{(i)}\pm d_i \in Q_i$. Consider the $r+1$ vectors of the form
	$A_id_i\in \R^r$. They have to be linearly dependent and thus, there
	exist $\lambda_1,\dots,\lambda_{r+1} ∈ℝ$ not all zero such that
	\[
	\sum_{i=1}^{r+1}\lambda_i A_i d_i=0
	\]
	By rescaling the $\lambda_i$ we can suppose that ${x^*}^{(i)}\pm \lambda_i d_i \in Q_i$. Consider
	\[
	d= (\lambda_1 d_1,\ldots,\lambda_{r+1} d_{r+1},0,\ldots,0)\in \R^{t_1+\dots+t_n} \setminus \{0\}
	\]
	Then $x^*+d, x^*-d$ are both feasible solutions of \eqref{eq:25} and $x^* = \frac{1}{2} (x^*+d) + \frac{1}{2} (x^*-d)$. So $x^*$ is a convex combination of two feasible points of \eqref{eq:25} and thus not a vertex.
\end{proof}

\begin{proposition}
\label{thr:4}
Let $F_i$ be the minimal face of the integral polyhedron $Q_i$ containing ${x^*}^{(i)}$. Then there exists an integer point ${y^*}^{(i)} ∈ F_i ∩ ℤ^{t_i}$ with
\begin{displaymath}
  \|{x^*}^{(i)} - {y^*}^{(i)}\|_1 ≤ r^2 G .
\end{displaymath}
\end{proposition}
The proof of this proposition uses standard arguments of polyhedral theory, see, e.g.~\cite{Schrijver86}. 
\begin{proof}
  The proposition is trivially true for $r=0$, henceforth we assume $r\geq 1$.  
  Let $w ∈F_i$ be an arbitrary vertex of $F_i$ and let $C_i$ be the cone
  \begin{displaymath}
    C_i = \{ λ ( f-w) ：f ∈ F_i, \, λ ≥0 \}. 
  \end{displaymath}
  The extreme rays of this cone are elements of the Graver basis of the matrix in a standard-form representation of $Q_i$, see, e.g.~\cite[Lemma 3.15]{onn2010nonlinear}. By assumption, each Graver basis element has $\ell_1$ norm bounded by G.
  The affine dimension of $F_i$ is bounded by $r$. 
  By Carathéodory's Theorem~\cite[p.~94]{Schrijver86} there exist $r$ Graver basis elements $g_1,\dots,g_r $ such that
  \begin{displaymath}
    {x^*}^{(i)}  = w  + ∑_{j=1}^r λ_j g_j.
  \end{displaymath}
  We consider the integer point $ z = w  + ∑_{j=1}^r ⌊λ_j⌋ g_j ∈ℤ^{t_i}$. There are two cases. If $z ∈ F_i$, then, by the triangle inequality,  the distance in $\ell_1$-norm of ${x^*}^{(i)}$ to the nearest integer point in $F_i$ is bounded by $r G$. Otherwise, the line segment between ${x^*}^{(i)} $ and $z$ exits $F_i$ in a lower dimensional face of $Q_i$ contained in $F_i$. 
  Call ${\tilde{x}}^{(i)}$ the intersection point. We apply an inductive argument now: there is an integer point on this lower dimensional face that has $\ell_1$-distance at most $(r-1)^2 G$ from ${\tilde{x}}^{(i)}$. The bound $r^2 G$ follows by applying the triangle inequality. 
\end{proof}

In the following, we keep the notation of Proposition~\ref{thr:4} and denote $({y^*}^{(1)},\dots,{y^*}^{(n)}) ∈ℤ^{t_1+\dots+t_n}$ by $y^*$. If $z^*∈ ℤ^{t_1+\dots+t_n}$ is a feasible integer solution of the block-structured IP~\eqref{eq:25} with integral $Q_i$, then each ${z^*}^{(i)}$ is in $Q_i$, in particular ${y^*}^{(i)} - {z^*}^{(i)}$ is in the kernel of the matrix in standard-form representation of $Q_i$.
Therefore, there exists a multiset $\sL_i$ of Graver basis elements of this matrix that are sign-compatible with ${y^*}^{(i)} - {z^*}^{(i)}$ such that
\begin{displaymath}
  {y^*}^{(i)} - {z^*}^{(i)} = ∑_{g ∈ \sL_i} g. 
\end{displaymath}

\begin{lemma}
  \label{lem:1}
  For each $i=1,\dots,n$ and each $\sH_i ⊆ \sL_i$ one has
  \begin{enumerate}[i)]
  \item \label{item:1} 
    \begin{align*}
      {z^*}^{(i)} + ∑_{h ∈\sH_i} h & \in Q_i \\
      {y^*}^{(i)} -∑_{h ∈\sH_i} h & \in Q_i. 
    \end{align*}
  \item There exists an $ε>0$ such that \label{item:2}
    \begin{align*}
      {x^*}^{(i)} - ε ∑_{h ∈\sH_i} h & ∈ Q_i.
    \end{align*}
  \end{enumerate}
\end{lemma}

\begin{proof}
  The Assertion~\ref{item:1}) follows from standard arguments (see e.g.~\cite{onn2010nonlinear}) as follows. Let $B_i$ be the matrix in a standard-form representation of $Q_i$. Then, one has
    \begin{align*}
      B_i \left({z^*}^{(i)} + \sum_{h \in \sH_i} h\right) & = b_i, \\
      B_i\left( {y^*}^{(i)} -\sum_{h \in \sH_i} h\right)  & = b_i
    \end{align*}
    and both ${z^*}^{(i)} + \sum_{h \in \sH_i} h$ and ${y^*}^{(i)} -∑_{h ∈\sH_i} h$ are integer. Since the Graver basis elements of $\sH_i$ are sign-compatible with ${y^*}^{(i)} - {z^*}^{(i)}$ the non-negativity is satisfied by both points as well. Thus both points are feasible integer points of the system $B_i x^{(i)}=b_i, \, x^{(i)} \ge 0$ which implies that they lie in $Q_i$.

    For the proof of Assertion~\ref{item:2}), 
    let the polyhedron $Q_i$ be described by the inequalities
    \begin{displaymath}
      Q_i = \{x ∈ ℝ^{t_i} ： D^{(i)}x ≤ p^{(i)}\} 
    \end{displaymath}
    for some integer matrix $D^{(i)} ∈ ℤ^{m_i × t_i}$ and integer vector $p^{(i)} ∈ℤ^{m_i}$. What is the inequality description of the minimal face  $F_i ⊆ Q_i$  containing  ${x^*}^{(i)}$?  Let  $I_i ⊆ \{ 1,\dots,m_i\}$ be the index set corresponding to the inequalities of $D^{(i)}x ≤ p^{(i)}$ that are satisfied by ${x^*}^{(i)}$ with equality. The inequality description of $F_i$ is obtained from  $D^{(i)}x ≤ p^{(i)}$ by setting the inequalities indexed by $I_i$ to equality.

    Since ${y^*}^{(i)} ∈ F_i$, all the inequalities indexed by $I_i$ and possibly more are also tight at ${y^*}^{(i)}$. However, subtracting $∑_{h ∈\sH_i} h$ from ${y^*}^{(i)}$, one obtains a point of $Q_i$. Therefore, we can move, starting at  ${x^*}^{(i)}$, in the direction of $-∑_{h ∈\sH_i} h$  some positive amount, without leaving $Q_i$. This means that assertion~\ref{item:2}) holds. 
\end{proof}

\begin{lemma}
  \label{lem:2}
  Suppose that $z^*$ is an optimal solution of the block-structured integer program~\eqref{eq:25} with integral polyhedra $Q_i$ that is closest to ${y^*}$ w.r.t. the $\ell_1$-distance. For each $i$, let $\sL_i$ be a multiset of Graver basis elements of the matrix in a standard-form representation of $Q_i$,  sign-compatible with ${y^*}^{(i)}- {z^*}^{(i)}$ such that ${y^*}^{(i)}-{z^*}^{(i)}$ decomposes into 
  \begin{displaymath}
    {y^*}^{(i)}-{z^*}^{(i)} = ∑_{h ∈ \sL_i} h, \quad i=1,\dots,n. 
  \end{displaymath}
  For each selection of sub-multisets $\sH_i \subseteq \sL_i$ for $i=1,\dots,n$
  one has
  \begin{displaymath}
    ∑_{i=1}^n ∑_{h ∈ \sH_i} A_i h ≠ 0. 
  \end{displaymath}
\end{lemma}

\begin{proof}
For the sake of contradiction, let $ \sH_1⊆ \sL_1, \cdots , \sH_n⊆ \sL_n$ be a selection of sub-multisets such that
\begin{equation}
\label{eq:21}
∑_{i=1}^n ∑_{h ∈ \sH_i} A_i h = 0 
\end{equation}
holds.  By Lemma~\ref{lem:1}~\ref{item:1}) one has ${z^*}^{(i)} + ∑_{h ∈\sH_i} h  ∈  Q_i$ for each $i$. Together, this implies that
\begin{equation}
\label{eq:22}
z^* + \left( ∑_{h ∈\sH_1}h, \dots, ∑_{h ∈\sH_n}h  \right) 
\end{equation}
is an integer 
solution.
Similarly, Lemma~\ref{lem:1}~\ref{item:2}) implies that there exists an $ε>0$ such that
\begin{displaymath}
x^* - ε \left( ∑_{h ∈\sH_1}h, \dots, ∑_{h ∈\sH_n}h  \right)  
\end{displaymath}
is a feasible solution of the relaxation of~\eqref{eq:25}. Since $z^*$ and $x^*$ were optimal solutions of the IP and the relaxation of~\eqref{eq:25} respectively, this implies that 
\begin{displaymath}
∑_{i=1}^n c_i^T ∑_{h ∈\sH_i}h = 0,
\end{displaymath}
and thus that the objective values of $z^*$ and~\eqref{eq:22} are the same. This however is a contradiction to the minimality of the $\ell_1$-distance of $z^*$ to $y^*$, as the optimal integer solution~\eqref{eq:22} is closer to $y^*$.
\end{proof}

As in the proximity result presented in~\cite{DBLP:journals/talg/EisenbrandW20}, we make use of the so-called  Steinitz lemma, which holds for an arbitrary norm $\|\cdot\|$.
\begin{theorem}[Steinitz (1913)]
	\label{thr:7}
	Let $x_1,\ldots,x_n \in \R^r$ such that 
	\begin{displaymath}
	\sum_{i=1}^n x_i = 0    \quad \text{ and } \quad  \|x_i\| \leq 1 \, \text{ for each }i. 
	\end{displaymath}
	There exists a permutation $π ∈ S_n$ such that all partial sums satisfy 
	$$\|\sum_{j=1}^k x_{\pi(j)}\| \leq c(r) \; \text{ for all } k=1,\ldots,n.$$
	Here $c(r)$ is a constant depending on $r$ only.
\end{theorem}

We now bound $\|y^* - z^*\|_1$ by using Theorem~\ref{thr:7} with  $c(r)=r$, see~\cite{grinberg1980value}.
The proof follows closely the ideas in~\cite{eisenbrand2018faster}.
\begin{theorem}
  \label{thr:5}
  Suppose that $z^*$ is an optimal solution of the block-structured integer program~\eqref{eq:25} that is closest to ${y^*}$ in $\ell_1$-norm.
  \begin{displaymath}
    \|y^* - z^*\|_1 ≤  (2r\Delta G+1)^{r + 4}. 
  \end{displaymath}
\end{theorem}
\begin{proof}
  We use the notation of the statement of Lemma~\ref{lem:2}. Denote the matrix $(A_1, \dots, A_n) ∈ℤ^{r × (t_1 +\dots+ t_n)}$ by $A$. We have
  \begin{align*}
  0 = A (x^* - z^*) & = A (x^* - y^* ) + A(y^* - z^*) \\
                    & = A (x^* - y^* ) + ∑_{i=1}^n ∑_{h ∈\sL_i} A_ih.                 
  \end{align*}
  Since the $\ell_1$-norm of each Graver basis element $h ∈ \sL_i$ is assumed to be bounded by $G$, the $\ell_∞$ norm of each $A_ih$ is bounded by $\Delta G$. 
  The Steinitz Lemma (Theorem \ref{thr:7}) implies that the set
  \begin{equation}
    \label{eq:24}
    \{A_i h \, : \, h \in \sL_i, \, i=1,\dots,n\}\subseteq \Z^r
  \end{equation}
  can be permuted in such a way such that the distance in the $\ell_∞$-norm of each prefix sum to the line-segment  spanned by $0$ and $A (x^* - y^*)$ is bounded by $\Delta G$ times the dimension $r$, i.e., by $R:=r \Delta G$. Note that each prefix sum is an integer point. The number of integer points that are within $\ell_\infty$-distance $R$ to the line segment spanned by $0$ and $A (x^* - y^*)$ is at most
  \begin{equation}
    \label{eq:23}
    (\|A (x^* - y^*)\|_1 +1) ⋅(2R+1)^r.  
  \end{equation}
  But
  \begin{align*}
    \|A (x^* - y^*)\|_1 & \leq \sum_{i=1}^n  \|A_i ({x^*}^{(i)} - {y^*}^{(i)})\|_1 \\
                       & \leq r ⋅ \Delta r^3 G,
  \end{align*}
  where the last inequality follows from Lemma~\ref{lem:bound-fractionality} and Proposition~\ref{thr:4}.
  Thus this number of integer points is bounded by
  \begin{displaymath}
    (r^4 \Delta G +1) (2r\Delta G+1)^r
  \end{displaymath}
  If $∑_{i=1}^n |\sL_i|$ is larger than this bound, then, in the Steinitz rearrangement of the vectors there exist two prefix sums, that are equal. This yields sub-multisets
\begin{displaymath}
    \sH_1⊆ \sL_1, \sH_2⊆ \sL_2, \cdots , \sH_n⊆ \sL_n
  \end{displaymath}
  for which one has 
  \begin{displaymath}
    ∑_{i=1}^n ∑_{h ∈ \sH_i} A_i h = 0. 
  \end{displaymath}
  By Lemma~\ref{lem:2}, this is not possible. This implies that
  \begin{displaymath}
    \|y^* - z^*\|_1 ≤ 
    (r^4 \Delta G +1) (2r\Delta G+1)^r = (r\Delta G)^{O(r)} . \qedhere
  \end{displaymath}
\end{proof}

Applying the triangle inequality $\|x^* - z^*\|_1 \leq \|x^* - y^*\|_1 +  \|y^* - z^*\|_1$, Theorem~\ref{thr:3} is proven.

\section{A dynamic program}
\label{sec:dynamic-program}

Let $x^*$ be an optimal vertex solution of a block-structured linear
programming problem where $Q_i$, $i=1,\dotsc,n$, are integral.
We now describe a dynamic programming approach that computes an optimal integer solution of the block-structured integer program.
We note that this approach closely resembles a method from~\cite{eisenbrand2019algorithmic}.

First we observe by Theorem~\ref{thr:3} that we can restrict our search to an
optimal integer solution $z$ with $\|x^* -z\|_1\le (r\Delta G)^{O(r)}$.
For simplicity of presentation we assume that $n$ is a power of $2$.
We construct a binary tree with leaves $1,\dotsc,n$.
Let $v_{j,k}$ be the $(j+1)$-th node on the $(k-1)$-th layer from the bottom.
Then $v_{j,k}$ corresponds
to an interval of the form $[j 2^k + 1, (j + 1) 2^{k}]$.
In particular, the root node corresponds to the interval $[1,n]$.
From the leaves to the root we compute solutions
$y^{(j 2^k + 1)}, \dotsc, y^{((j + 1) 2^k)}$ for each $v_{j,k}$
using the solutions of the two children of the current node.

For some integer vector $z$ with the proximity above one has in the linking constraints for each $v_{j,k}$ 
\begin{equation}
  \label{eq:1}
  \left\| \sum_{i=j 2^k + 1}^{(j + 1)2^k} A_i \left( {x^*}^{(i)} -z^{(i)} \right) \right\|_\infty \le (r\Delta G)^{O(r)},
\end{equation}
which implies the following bounds
\begin{equation}
  \label{eq:3}
    \sum_{i=j 2^k + 1}^{(j + 1)2^k} A_i {x^*}^{(i)} -  (r\Delta G)^{O(r)}\mathbf{1} \le \sum_{i=j 2^k + 1}^{(j + 1)2^k} A_i z^{(i)} \le \sum_{i=j 2^k + 1}^{(j + 1)2^k} A_i {x^*}^{(i)} + (r\Delta G)^{O(r)}\mathbf{1},
\end{equation}
where $\mathbf{1}$ means the all-ones vector.
Let $S_{j,k} \subseteq \Z^r$ be the set of integer vectors $d \in \Z^r$ that satisfy 
\begin{equation}
  \label{eq:4}
  \sum_{i=j 2^k + 1}^{(j + 1)2^k} A_i {x^*}^{(i)} - (r\Delta G)^{O(r)}\mathbf{1} \le d \le \sum_{i=j 2^k + 1}^{(j + 1)2^k} A_i {x^*}^{(i)} + (r\Delta G)^{O(r)}\mathbf{1}. 
\end{equation}
We generate all these sets $S_{j,k}$. Clearly, the cardinality of each $S_{j,k}$ satisfies
\begin{equation}
  \label{eq:18}
  |S_{j,k}| \le (r\Delta G)^{O(r^2)}. 
\end{equation}
Now we compute for each node $v_{j,k}$ and each $d\in S_{j,k}$ a partial solution
$y^{(j 2^k + 1)}, \dotsc, y^{((j + 1) 2^k)}$ bottom-up.
The solution $y$ will satify the following condition: If $d$ is the the correct guess, that is to say,
\begin{equation}
\sum_{i=j 2^k + 1}^{(j + 1)2^k} A_i z^{(i)} = d ,
\end{equation}
then our computed partial solution $y$ satisfies
\begin{equation}
\sum_{i=j 2^k + 1}^{(j + 1)2^k} c^T_i y^{(i)} \ge \sum_{i=j 2^k + 1}^{(j + 1)2^k} c^T_i z^{(i)} .
\end{equation}
For the leaves we simply solve the individual block using a presumed algorithm we have for them, that is, we compute
the optimal solution to
\begin{equation}
  \label{eq:19}
  \max\left\{ c_i^Tx^{(i)} ： A_i x^{(i)} = d, \, x^{(i)} \in Q_i, \, x^{(i)} \in \Z^{t_i}_{\ge 0} \right\} .
\end{equation}
For an inner node $v_{j,k}$ with children $v_{2j, k-1}$ and $v_{2j+1, k-1}$ we consider all 
$d'\in S_{2j, k-1}$ and $d''\in S_{2j+1, k-1}$ with $d = d' + d''$ and take the best solution among
all combinations. Indeed, if $d$ is the correct guess, then the correct guesses $d'$ and $d''$
are among these candidates.

After computing all solutions for the root, we obtain an optimal solution by taking the solution for
$b_0 \in S_{0, \log(n)}$.

The algorithm performs $\log(n)$ rounds, where the first one consists of
running the algorithm for the block problems~\eqref{eq:19} and all others in
computing maxima over $(r\Delta G)^{O(r^2)}$ elements. This maximum
computation can be done in constant time on a number of processors
that is quadratic in the number of elements, see~\cite{DBLP:books/daglib/0031448}.
\begin{theorem}
	\label{thr:round-parallel}
	There is a linear algorithm that rounds a vertex solution $x^*$
        of the block-structured linear programming problem with integral $Q_i$,$i=1,\dotsc,n$, to an integral one using
	\[O(\log(n)+T_{\max})\]
	operations on each of $(r \Delta G)^{O(r^2)}R$ processors. Here $R=\sum_{i}R_i$ is the sum of the processor requirements of the algorithms for the block problems~\eqref{eq:19} and $T_{\max}$ is the maximum number of operations of any of them. 
\end{theorem}

\section{Applications}
\label{sec:applications}

In this section we derive running time bounds for concrete cases of block-structured integer programming problems using the theorems from the earlier sections. All algorithms in this section are linear algorithms in the sense of the definition in Section~\ref{sec:prelim}.
First we need to bound the increase in the norm of Graver elements when linking blocks with $r$ additional constraints.
\begin{proposition}[Eisenbrand et al.~\cite{eisenbrand2018faster}]
  \label{prop:graver}
  Consider the block-structured problem \eqref{eq:25} where the $Q_i$ are integer hulls of $\{x^{(i)} : B_i x^{(i)}, x^{(i)} \geq 0\}$ for some matrices $B_i$.
  Suppose the $\ell_1$-norm of the Graver elements of matrices $B_1,\dotsc,B_n$ is bounded by $G$.
  Then the $\ell_1$-norm of the Graver elements of the matrix representing~\eqref{eq:25} is
  bounded by 
\begin{equation}
  G \cdot (2rG\Delta + 1)^{r} .
\end{equation}
\end{proposition}

Our base case is the trivial integer program
\begin{equation}\label{eq:base-case}
  \max\{cx : A x = b, 0\le x\le u, x \in \Z\} ,
\end{equation}
where $c\in\R, A\in \Z^{s\times 1}$, $b\in\Z^s$,
and $u\in\Z\cup\{\infty\}$.
Clearly this problem can be solved in $O(s)$ operations.

\begin{corollary}
\label{cor:ew}
Let $c\in\R^t, A\in \Z^{s\times t}$, $b\in\Z^m$, and $\Delta \ge \lVert A \rVert_\infty$.
Consider the integer programming problem
\begin{equation}
  \max\{c^T x : A x = b, 0\le x\le u, x \in \Z^t\} .\label{eq:ew-case}
\end{equation}
This problem can be solved in
\begin{equation*}
  (s\Delta)^{O(s^2)} t^{1 + o(1)}
\end{equation*}
arithmetic operations, or alternatively, in $2^{O(s^2)} \log^{s+1}(t)$ operations on $(s\Delta)^{O(s^2)} t$ processors in the PRAM model.
\end{corollary}
\begin{proof}
Using Theorem~\ref{thr:lp-parallel} with $Q_i = \R$ for all $i$, we can solve the relaxation in $2^{O(s^2)}\log^{s+1}(t)$ operations on $t$ processors.
Then, with an upper bound of $O(\Delta)$ on the $\ell_1$-norm of a Graver element of \eqref{eq:base-case} we use Theorem~\ref{thr:round-parallel} to derive an optimal integer solution in $O(\log(st))$ operations on $(s\Delta)^{O(s^2)} t$ processors.

The sequential running time follows with the fact that $\log^k(n) \le k^{2k} n^{o(1)}$ for all $k$.
\end{proof}
\begin{corollary}\label{cor:nfold}
Let $c\in\R^{nt}, u\in\Z^{nt}_{\ge 0}, b_0\in \Z^r$ and for all $i=1,\dotsc,n$ let
$A_i \in \Z^{r\times t}$, $B_i \in \Z^{s \times t}$, and $b_i \in \Z^{s}$.
Furthermore, let $\Delta \ge \lVert A_i \rVert_\infty, \lVert B_i \rVert_\infty$
for all $i=1\dotsc,n$.
Consider the integer programming problem
\begin{align}
  \max &\ c^T x \notag\\
  \begin{pmatrix}
    A_1 & \dotsc & A_n \\
    B_1 &        &     \\
        & \ddots &     \\
        &        & B_n
  \end{pmatrix}
  x &=
  \begin{pmatrix}
    b_0 \\
    b_1 \\
    \vdots \\
    b_n
  \end{pmatrix} \label{eq:nfold-case} \\
  0 \le x &\le u \notag\\
  x &\in \Z^{nt} .\notag
\end{align}
This problem can be solved in
\begin{equation*}
  2^{O(rs^2)}(rs\Delta)^{O(r^2 s + s^2)} (nt)^{1 + o(1)}
\end{equation*}
arithmetic operations, or alternatively, in $2^{O(r^2 + rs^2)} \log^{O(rs)}(nt\Delta)$ operations on $(rs\Delta)^{O(r^2 s + s^2)} nt$ processors in the PRAM model.
\end{corollary}

\begin{proof}
Using Theorem~\ref{thr:lp-parallel} we can solve the relaxation
in $2^{O(r^2 + r s^2)}\log^{O(rs)}(t)\log^{r+1}(nt\Delta)$ operations on $(s\Delta)^{O(s^2)}nt$ processors.
From Proposition~\ref{prop:graver} we derive that a Graver element
of~\eqref{eq:ew-case} has an $\ell_1$-norm of at most $(s\Delta)^{O(s)}$.
Then using Theorem~\ref{thr:round-parallel} we derive an optimal integer solution
in $O(2^{O(r+s)^2} \log^{r+s+1}(t)+\log(n))$ operations on $(rs\Delta)^{O(r^2 s + s^2)} nt$ processors.
\end{proof}
Another parameter under consideration for integer programming is the (dual) treedepth of the constraint matrix $A$.
As we are only interested in giving a comparison to other algorithms in the literature, we omit a thorough discussion.
See e.g.~\cite{koutecky2018parameterized} for the necessary framework.
Using this framework, a clean recursive definition of dual treedepth,
can be written as follows.
The empty matrix has a treedepth of $0$.
A matrix $A$ with treedepth $d > 0$ is either block-diagonal and the maximal treedepth of the blocks is $d$, or there is a row such that after its deletion, the resulting matrix has treedepth $d-1$.
The necessary structure, i.e.\ which row of $A$ should be deleted, can be determined via some preprocessing whose running time is negligible. For simplicity we assume this knowledge available to us.
It is known, and can be proven with Proposition~\ref{prop:graver}, that the Graver basis elements of a matrix with treedepth $d$ have $\ell_1$-norm at most $(3\Delta)^{2^d - 1}$~\cite{knop2019tight}.
Though sometimes the running time is analyzed in a more fine-grained way
by introducing additional parameters,
the best running time in literature purely on parameter $d$ and $\Delta$
have a parameter dependency of $\Delta^{O(d2^d)}$.\cite{knop2019tight}

\begin{restatable}{corollary}{treedepthIP}
An integer programming problem $\max\{c^T x : Ax = b, x\in\Z^h\}$,
where $A$ has dual treedepth $d$ and $\Delta \ge \lVert A \rVert_\infty$
can be solved in
\begin{equation*}
2^{O(d 2^d)} \Delta^{O(2^d)} h^{1 + o(1)}
\end{equation*}
arithmetic operations, or alternatively, $2^{O(d2^d)}\log^{O(2^d)}(\Delta h)$ parallel operations on $\Delta^{O(2^d)}h$ processors.
\end{restatable}

The proof requires tedious calculations and is deferred to the appendix.

\subsection*{Continuous variables}
We now consider cases of block-structured linear programming,
i.e., the domain of the variables is $\R_{\ge 0}$.
Although linear programming has polynomial algorithms, no
strongly polynomial algorithm is known for the general case and there is
 no PRAM algorithm running in polylogarithmic time on a polynomial number of processors unless $\mathrm{NC} = \mathrm{P}$~\cite{greenlaw1995limits}.
For the following corollaries, we only use Theorem~\ref{thr:lp-parallel}.

The continuous variant of the base case~\eqref{eq:base-case} is easily solvable
in $O(s)$ operations.
\begin{corollary}
\label{cor:standard-LP}
  The continuous variant of~\eqref{eq:ew-case} can be solved
  in
\begin{equation*}
  2^{O(s^2)} t^{1 + o(1)}
\end{equation*}
  operations, or alternatively,
  in $2^{O(s^2)}\log^s(t)$ operations on each of $t$ processors in the PRAM model.
\end{corollary}
\begin{corollary}
  The continuous variant of~\eqref{eq:nfold-case} can be solved
  in
\begin{equation*}
  2^{O(r^2 + r s^2)} (nt)^{1 + o(1)}
\end{equation*}
  operations, or alternatively,
  in $2^{O(r^2 + r s^2)}\log^{O(rs)}(nt)$ operations on each of $nt$ processors in the PRAM model.
\end{corollary}

\begin{corollary}
Let $A \in \Z^{m \times h}$ be a matrix with treedepth $d$.
Then the LP $\max\{c^T x : \ Ax = b, \ x \geq 0\}$ can be solved in
\begin{equation*}
2^{O(d2^d)}h^{1+o(1)}
\end{equation*}
operations, or alternatively, with $2^{O(2^d)}\log^{2^{d+1}-2}(h)$ operations on each of $h$ processors in the PRAM model.
\end{corollary}
\begin{proof}
We induct on $d$. For $d=1$, $A$ consists of a single constraint and Corollary \ref{cor:standard-LP} for $s=1$ gives a running time of $2^{O(1)}\log(h)$ on each of $h$ processors.
For $d \ge 2$, the matrix $A$ has a block-structure with $r=1$ and all polyhedra $Q_i$ of dual treedepth at most $d-1$. Using the recursion hypothesis and Theorem \ref{thr:lp-parallel} we get the desired result. 
\end{proof}



\appendix
\section*{Appendix}

\megiddo*

\begin{proof} 
  We show that in $2^{O(r)}\log(m)$ operations on $O(m)$ processors with
  $2^{O(r)}$ comparisons dependent on $\bar\lambda$ we can determine
  the location of $\bar\lambda$ with respect to half of the hyperplanes.
  Then with $O(\log(m))$ repetitions the proposition follows.
  The algorithm solves the problem by recursing to the same problem in smaller dimensions.
  
  Consider the base case $r = 1$.
  We first consider the hyperplanes $i$ with $a_i = 0$. Here we need to check whether 
   $\bar{f}_i \le 0$ and  $\bar{f}_i \ge 0$ which can be done in parallel. 
  Hence, assume that $a_i \neq 0$ for all $i$.  
  Next we compute the median  $\bar M$  of the numbers $\bar{f}_i / a_i$
  and for each hyperplane $H_i$ whether $\bar{f}_i / a_i \le \bar M$ and $\bar{f}_i / a_i \ge \bar M$ hold.
  The median computation requires $O(\log(m))$ operations on $O(m)$ processors,
  for example by a straightforward implementation of merge sort, see~\cite{DBLP:books/daglib/0031448}.
  Now compare $\bar{\lambda} \le \bar M$ and $\bar{\lambda} \ge \bar M$. From the result we can derive an answer
  for $m/2$ of the hyperplanes. The total number of linear queries involving $\bar{λ}$  is bounded by $O(\log m)$. 

  Now let $r > 1$. In this case the crucial observation is that when we have one hyperplane $H_i$ with
  $(a_i)_1 / (a_i)_2 \le 0$ and another hyperplane $H_j$ with $(a_j)_1 / (a_j)_2 > 0$, we can define two
  new hyperplanes -- one that is parallel to the $\lambda_1$-axis, and one that is parallel to $\lambda_2$-axis --
  such that by locating $\bar\lambda$ with respect to the new hyperplanes we can derive the location with
  respect to one of $H_i$ and $H_j$. Towards this we first transform the coordinate system so that many pairs with 
  this property exist.
  We take all hyperplanes $H_i$ with $(a_i)_2 = 0$ and solve half of them recursively
  in dimension $r-1$.
  For the remainder assume w.l.o.g.\
  that $(a_i)_2 > 0$ for all $H_i$ by swapping the signs on other hyperplanes.
  We now compute the median $M$ of $(a_i)_1 / (a_i)_2$ and for each hyperplane $H_i$ determine whether $(a_i)_1 / (a_i)_2 \le M$
  and $(a_i)_1 / (a_i)_2 \ge M$ hold.
  This takes $O(\log(m))$ operations on $O(m)$ processors.

  Next we transform the coordinate system using the automorphism defined by $F\in\R^{r\times r}$
  with
  \begin{equation*}
    (\lambda_1, \lambda_2, \lambda_3, \cdots, \lambda_r)^T F
    = (\lambda_1 - M \lambda_2, \lambda_2, \lambda_3, \cdots, \lambda_r)^T
  \end{equation*}
  For each $H_i$ define
  a new hyperplane $H'_i = \{\lambda' \in \R^r : a_i^{\prime T} \lambda' = \bar{f}_i\}$ with $a' = a F$.
  Locating $\bar{\lambda'} = (F^T)^{-1}\bar{\lambda}$ with respect to the hyperplanes $H'_i$ is equivalent
  to locating $\lambda$ with respect to the hyperplanes $H_i$.
  Hence, it suffices to construct an algorithm for the new hyperplanes $H'_i$.
  Then we run this algorithm, but transform each comparison $u^T \bar{\lambda'} + v^T \bar{f} \le w$ to $(u^T (F^T)^{-1}) \bar{\lambda'} + v^T \bar{f}\le w$.

  We have established that for half of the new hyperplanes $(a'_i)_1 \le 0$ holds
  and $(a'_i)_1 \ge 0$ for the other half. Notice also that $(a'_i)_2 = (a_i)_2 > 0$.
  We form a maximum number of pairs of hyperplanes such that
  for each pair $(H'_i, H'_j)$ we have
  $(a'_i)_1 \le 0$ and $(a'_j)_1 > 0$.
  For all remaining hyperplanes we have $(a'_i)_1 = 0$
  and we solve half of these by recursing to $r-1$. Hence, we focus on the pairs.
  We alter the hyperplanes once more. Define for each pair $(H'_i, H'_j)$ the two hyperplanes
  \begin{align*}
  H''_i &= \bigg\{\lambda'\in\R^r : \big(\underbrace{(a'_j)_1 a'_i - (a'_i)_1 a'_j}_{=: a''_i}\big)^T \lambda' = \underbrace{(a'_j)_1 \bar{f}_i - (a'_i)_1 \bar{f}_j}_{=: \bar{f''}_i} \bigg\} , \\
  H''_j &= \bigg\{\lambda'\in\R^r : \big(\underbrace{(a'_j)_2 a'_i - (a'_i)_2 a'_j}_{=: a''_i}\big)^T \lambda' = \underbrace{(a'_j)_2 \bar{f}_i - (a'_j)_1 \bar{f}_j}_{=: \bar{f''}_j} \bigg\} .
  \end{align*}
  These are combinations that eliminate either the first coordinate or the second.
  Hence, locating $\bar{\lambda'}$ with respect to the hyperplanes of the first kind (those where $\lambda_1$ is eliminated)
  is a problem in dimension $r-1$. We solve recursively half of these hyperplanes.
  Then we recurse again on the hyperplanes of the second kind,
  but only for those pairs where we already solved the first hyperplane.
  Thus, we have located $\bar{\lambda'}$ with respect to both $H''_i$ and $H''_j$ for
  at least $1/4$ of the pairs $(H'_i, H'_j)$.
  It remains to derive the location with respect to one of $H'_i$ and $H'_j$.
  In two dimensions this is illustrated in Figure~\ref{fig:megiddo}.
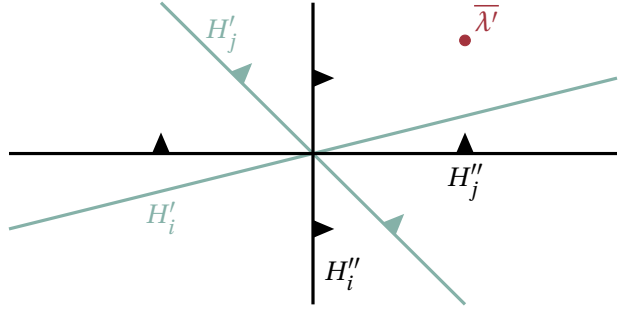
\begin{figure}
\begin{center}
\begin{tikzpicture}
  \draw[very thick, mygreen] (-4, -1) -- (4, 1) node[pos=0.25, below] {$H'_i$};
  \draw[very thick, mygreen] (-2, 2) -- (2, -2) node[pos=0.2, above] {$H'_j$};
  \draw[thick, mygreen, -triangle 45] (-1, 1) -- (-1 + 0.2, 1 + 0.2);
  \draw[thick, mygreen, -triangle 45] (1, -1) -- (1 + 0.2, -1 + 0.2);
  \draw[very thick] (-4, 0) -- (4, 0) node[pos=0.75, below] {$H''_j$};
  \draw[thick, -triangle 45] (-2, 0) -- (-2, 0.28);
  \draw[thick, -triangle 45] (2, 0) -- (2, 0.28);
  \draw[very thick] (0, -2) -- (0, 2) node[pos=0.1, right] {$H''_i$};
  \draw[thick, -triangle 45] (0, 1) -- (0.28, 1);
  \draw[thick, -triangle 45] (0, -1) -- (0.28, -1);
  \draw[fill, myred] (2, 1.5) circle (2pt) node[above right] {$\bar{\lambda'}$};
\end{tikzpicture}
\end{center}
\caption{Megiddo's multidimensional search algorithm illustrated in $r=2$ dimensions}
\label{fig:megiddo}
\end{figure}
  Notice that we can write
  \begin{align*}
    \underbrace{\left((a'_i)_2 (a'_j)_1 - (a'_i)_1 (a'_j)_2 \right)}_{:= \mu_i > 0} a'_i = \underbrace{(a'_i)_2}_{\ge 0} a''_i - \underbrace{(a'_i)_1}_{\le 0} a''_j , \quad (a'_i)_2 \bar{f''}_i - (a'_i)_1 \bar{f''}_j = \mu_i \bar{f}_i \\
    \underbrace{\left((a'_j)_1 (a'_i)_2 - (a'_j)_2 (a'_i)_1 \right)}_{:= \mu_j < 0} a'_j = \underbrace{(a'_j)_2}_{\ge 0} a''_i - \underbrace{(a'_j)_1}_{> 0} a''_j , \quad (a'_j)_2 \bar{f''}_i - (a'_j)_1 \bar{f''}_j = \mu_j \bar{f}_j  .
  \end{align*}
  From the coefficients above it follows that if the signs of the comparisons
  between $a^{\prime\prime T}_i \bar{\lambda'}, \bar{f''}_i$ and
  $a^{\prime\prime T}_j \bar{\lambda'}, \bar{f''}_j$
  are equal, this implies an answer for $H'_j$ and otherwise it implies an answer for $H''_i$.
  This means we have solved $1/8$ of these hyperplanes in the pairs.
  After repeating this procedure a constant number of times,
  we know the location to at least half of the hyperplanes.
  This procedure requires $O(\log(m))$ operations on $O(m)$ processors and no comparisons (except for those made by recursions).
  The number of recursive calls to dimension $r-1$ is also constant. This leads to a total running time (including recursions)
  of $2^{O(r)}\log(m)$ on $O(m)$ processors and $2^{O(r)}$ comparisons on $\bar{\lambda}$. We note that
  the arithmetics on $\bar{f}_i$ increase in the recursions, since we have to make these operations on linear functions
  in $\bar{f}_i$. However, the cardinality of the support of these functions is always bounded by $2^{O(r)}$.
  Hence, this overhead is negligible.
\end{proof}

\lprelbad*

\begin{proof}
We construct a family of problems in which the number $n$ of blocks is odd, the number of variables is two for each block.
We denote them by $x^{(i)}_1$ and $x^{(i)}_2$, $1≤i≤n$ and they are constrained to be non-negative.

We maximize
\[
\sum_{i=1}^n (2+ε) x^{(i)}_{1} + (3-ε) x^{(i)}_2 
\]
under the blocks given by the polyhedra 
\[Q_i=\{x^{(i)} : 2 x^{(i)}_{1} + 3 x^{(i)}_{2} = 3, x^{(i)} \geq 0\},  i=1,\dots,n-1 \]
and
\[Q_n=\{x^{(n)} : 2 x^{(n)}_{1} + 3 x^{(n)}_{2} = 6n\}.\]

Notice that for these constraints, the unique optimal LP-solution is given by ${x^*_1}^{(i)} = 3/2$ and ${x^*_{2}}^{(i)} = 0$ for  $i=1,\dots,n-1$ and ${x^*_{1}}^{(n)} = 3n$ and ${x^*_{2}}^{(n)} = 0$ respectively. 
We add the linking constraint
\begin{displaymath}
\sum_{i=1}^{n-1} (x^{(i)}_1 +  x^{(i)}_2 ) - (x^{(n)}_1 + x^{(n)}_2) = 3/2(n-1) - 3n
\end{displaymath}
which is satisfied by the optimal LP solution $x^*$.
It may be checked that the optimal IP-solution is defined by setting
${z^*_{1}}^{(i)}  = 0, \,{z^*_{2}}^{(i)} = 1, \, i=1,\dots,n-1$, 
${z^*_{1}}^{(n)} =  3 n - 3(n-1)/2$ and ${z^*_{2}}^{(n)} = n-1$. We obtain that ${z^*_{2}}^{(n)}-{x^*_{2}}^{(n)} = n-1$ giving the lower bound of Proposition \ref{prop:1}.
\end{proof}

%

\treedepthIP*

\begin{proof}
	We show the Corollary by induction on the treedepth for a slightly more general problem.
	Let $\binom{A}{B}$ be a constraint matrix with $A \in \Z^{r \times h}$, $r \ge 1$, $B \in \Z^{m \times h}$, and $\lVert A \rVert_\infty, \lVert B \rVert_\infty\le \Delta$. Moreover, assume that $B$ has a treedepth of $d$.
	Our induction hypothesis is that 
	this problem can be solved in $T(d, r, h)$
	operations on each of $R(d, r, h)$ processors, where
	\begin{equation*}
	R(d, r, h) := (r + d)^{c(d+1)(r+d)^2} (3\Delta)^{c 2^d (r + 4)^2} h
         = (r\Delta)^{O(2^d r^2)} h
	\end{equation*}
        and
	\begin{equation*}
	T(d, r, h) := 2^{c(2^{d+1} - 1)(r + 4)^3 }\log^{ (2^{d+1} - 1)(r + 1)}(R(d,1,h)) = 2^{O((d+r^3) 2^d)} \log^{O(2^d r)}(\Delta h) .
	\end{equation*}
	Here $c$ is a sufficiently large constant.
	In particular, we assume that $c$ is much larger than
	any hidden constant appearing in the $O$-notation of the Theorems~\ref{thr:round-parallel} and~\ref{thr:lp-parallel}.
        The requirement $r\ge 1$ is only to avoid corner cases in the proof.
        Although we are ultimately interested in $r = 0$, i.e., matrix $A$ is omitted,
        we can simply apply the theorem with $r=1$ and use a trivial zero matrix $A$.

	For $d=0$ and any $r \geq 1$, we rely on Corollary~\ref{cor:ew}.
	Now suppose that $d \geq 1$, and let us outline the core of the induction step.
	We can assume that $B$ is block-diagonal with $n \geq 1$ blocks $B_i$ that do not decompose any further.
	We split $A = (A_1,\dots, A_n)$ accordingly and
        we are interested in the subproblems $\binom{A_i}{B_i}$.
	However, for each $i$, this matrix can also be considered as $\binom{A_i'}{B_i'}$ where $A_i'$ has $r+1$ rows and $B_i'$ has treedepth at most $d-1$.
	We write $t_i$ for the number of variables in the $i$-th subproblem.
	First, we analyze the processor usage.
	For solving the LP we require
	\begin{equation*}
	\sum_{i=1}^n R(d-1,1,t_i) \le d^{cd^3} (3\Delta)^{c 2^{d-1} (1 + 4)^2} (t_1 + \cdots + t_n)
	\le (r + d)^{c (d + 1)(r + d)^2} (3\Delta)^{c 2^{d} (r + 4)^2} h
	\end{equation*}
	processors.
	For rounding to an integer solution, we require at most
	\begin{align*}
	(r\Delta G)^{c/2\cdot r^2} \sum_{i=1}^n R(d-1,r+1,t_i)
	&\le 
	r^{c r^2} (3\Delta)^{c/2\cdot 2^d r^2}
	\sum_{i=1}^n R(d-1,r+1,t_i) \\
	&\le 
	r^{c r^2} (3\Delta)^{c 2^{d-1} r^2} \cdot
	(r + d)^{cd(r + d)^2} (3\Delta)^{c 2^{d-1} (r + 1 + 4)^2} (t_1 + \cdots + t_n) \\
	&\le 
	(r + d)^{c (d + 1)(r + d)^2} (3\Delta)^{c 2^{d} (r + 4)^2} h
	\end{align*}
	processors.
	For the last inequality notice that
	$2 (r + 4)^2 \ge (r + 1 + 4)^2 + r^2$ for all $r\geq 0$.
	This proves the closed formula for the number of processors.
	
	In order to establish a recursive formula for the running time, we consider the running time for solving the strengthened relaxation and for rounding.
	We obtain
	\begin{align*}
	T(d,r,h) &\le 1/2 \cdot 2^{cr^2}(T(d-1,1,h) \cdot \log(R(d-1,1,h)))^{r+1} \\
                 &\quad + 1/4 \cdot 2^c (\log(n) + T(d-1,r+1,h)) .
	\end{align*}
        Here we use that $\sum_{i=1}^n R(d-1,1,t_i) = R(d-1,1,h)$.
	To bound the first summand we calculate
	\begin{align*}
	  & 1/2 \cdot 2^{cr^2}\left(T(d-1,1,h) \cdot \log(R(d-1,1,h))\right)^{r+1} \\
	&\le 1/2\cdot 2^{c r^2} \left(2^{125 c (2^{d} - 1)}\log^{ 2(2^{d} - 1)}(R(d-1,1,h)) \cdot \log(R(d-1,1,h))\right)^{r+1} \\
	&\le 1/2\cdot 2^{c r^2 + 125c (2^{d} - 1)(r+1)}\log^{ (2(2^{d} - 1) + 1)(r + 1)}(R(d-1,1,h)) \\
	&\le 1/2\cdot 2^{c (r+4)^3 + 2c (2^{d} - 1)(r+4)^3}\log^{ (2^{d+1} - 1)(r + 1)}(R(d-1,1,h))\\
	&\le 1/2\cdot 2^{c (2^{d+1} - 1)(r+4)^3}\log^{ (2^{d+1} - 1)(r + 1)}(R(d,1,h)) .
	\end{align*}
	Moreover, for the second summand we get
	\begin{align*}
	1/4 \cdot 2^c (\log(n) + T(d-1, r+1, h))
	&\le 1/2 \cdot 2^c T(d-1, r+1, h) \\
	&\le 1/2 \cdot 2^c \cdot 2^{c(2^{d} - 1)(r + 5)^3}\log^{(2^{d} - 1)(r + 2)}(R(d-1, 1, h)) \\
	&\le 1/2 \cdot2^{c(2^{d+1} - 1)(r + 4)^3 }\log^{(2^{d+1} - 1)(r + 1)}(R(d,1,h)) .
	\end{align*}
	Adding the two bounds concludes the proof of the running time.
\end{proof}

\end{document}